\documentclass[twocolumn,twoside]{IEEEtran}

\ifCLASSINFOpdf

\else

\fi

\ifCLASSOPTIONcompsoc
\usepackage[caption=false, font=normalsize, labelfont=sf, textfont=sf]{subfig}
\else
\usepackage[caption=false, font=normalsize]{subfig}
\fi
\usepackage{lipsum}%
\usepackage[dvipsnames]{xcolor}

\usepackage{balance}
\usepackage{multicol}   % for equalize of last page columns
\usepackage{cite}
\usepackage{gensymb}
\usepackage{multirow}
\usepackage{graphics}
\usepackage{epsfig}
\usepackage{graphicx}
\usepackage{epstopdf}
\usepackage{textcomp}
\usepackage{amsmath}
\usepackage{mathtools}
\usepackage{filecontents}
\usepackage{lipsum,color}
\usepackage{amssymb}
\usepackage{float}
\usepackage{comment}
\usepackage{amsthm}

\usepackage{braket}
\usepackage[T1]{fontenc}
\usepackage{mathptmx}   

\usepackage{amsthm}  % for proof 
\usepackage{amsfonts}

\theoremstyle{break}
\theoremstyle{break}
\begin{document}
\title{Deep Learning Surrogate for Fast CIR Prediction in Reactive Molecular Diffusion–Advection Channels}

\author{Meysam~Ghanbari,~Mohammad~Taghi~Dabiri,
	~Mazen~Hasna,~{\it Senior Member,~IEEE},\\
	~Tanvir~Alam,
	~and~Khalid~Qaraqe,~{\it Senior Member,~IEEE}
\thanks{M. Ghanbari, M.T. Dabiri, and Tanvir Alam are with the College of Science and Engineering, Hamad Bin Khalifa University, Doha, Qatar (Email: megh89467@hbku.edu.qa; mdabiri@hbku.edu.qa; talam@hbku.edu.qa).}

	\thanks{Mazen Hasna is with the Department of Electrical Engineering, Qatar University, Doha, Qatar (Email: hasna@qu.edu.qa).}
	\thanks{Khalid A. Qaraqe is with the College of Science and Engineering, Hamad Bin Khalifa University, Doha, Qatar, and also with the Department of Electrical Engineering, Texas A\&M University at Qatar, Doha, Qatar (Email: kqaraqe@hbku.edu.qa).}
	\thanks{This publication was made possible by Texas A\&M University at Qatar and Hamad Bin Khalifa University, which supported this publication. } %The statements made herein are solely the responsibility of the author[s].

}

% make the title area
\maketitle
%\vspace{-1cm}
%%%%%%%%%%%%%%%%%%%%%%%%%%%%%%%%%%%%%%%%%%%%%%%%%%%%%%%%%%
%%%%%%%%%%%%%%%%%%%%%%%%%%%%%%%%%%%%%%%%%%%%%%%%%%%%%%%%%%
\begin{abstract}
Accurate channel impulse response (CIR) modeling in molecular communication (MC) often requires solving coupled reactive diffusion–advection equations, which is computationally expensive for large parameter sweeps or design loops. We develop a deep-learning surrogate for a three-dimensional duct MC channel with reactive diffusion–advection transport and reversible ligand–receptor binding on a finite ring receiver. Using a physics-based partial differential equation (PDE)–ordinary differential equation (ODE) model, we generate a large CIR dataset across broad transport, reaction, and geometric ranges and train a neural network that maps these parameters directly to the CIR. On an independent test set, the surrogate closely matches reference CIRs both qualitatively and quantitatively: the empirical cumulative distribution function (CDF) of the normalized root mean square error (NRMSE) shows that 90\% of test channels are predicted with error below ~0.15, with only weak dependence on individual parameters. The surrogate therefore offers an accurate and computationally efficient replacement for repeated PDE-based CIR evaluations in MC system analysis and design.
\end{abstract}

\begin{IEEEkeywords}
Molecular Communication, Reactive Ducts, Channel Impulse Response, Deep Learning Surrogates
\end{IEEEkeywords}

\IEEEpeerreviewmaketitle

%% ---------------------------------------
%% ---------------------------------------
\section{introduction}
Molecular communication (MC) is an emerging paradigm for nano–bio interfaces and microfluidic platforms, where information is carried by signaling molecules rather than electromagnetic waves. In a typical MC link, a transmitter releases molecules into a medium where they diffuse, experience flow, and may degrade before reaching a receiver with reactive or receptor-coated surfaces \cite{Hamidovic2024Microfluidic}. These transport and reaction processes determine how the transmitted signal is distorted. The channel impulse response (CIR) captures their combined effect and underpins modulation, detection, synchronization, and performance analysis. Accurate and efficient CIR prediction in realistic geometries is therefore crucial for practical MC system design. However, for most realistic MC channels, no analytical closed-form expression for the CIR is available. Instead, the CIR must usually be computed by numerically solving a coupled 3D advection–diffusion–reaction partial differential equation (PDE) together with surface ordinary differential equations (ODE) that describe receptor binding and unbinding at the receiver. These models are often stiff and computationally demanding, particularly when many CIR evaluations are required for parameter sweeps, system optimization, or adaptive link operation in real time \cite{Jamali2019Tutorial}.

The authors in \cite{Bhatnagar2025Heterogeneous} studied heterogeneity and mobility in an unbounded 3D purely diffusive molecular channel, where the diffusivity of nodes and molecules varies randomly in time, leading to a non-Gaussian, time-varying CIR. In \cite{Oner2023SphericalShell}, a bounded but still diffusion-driven spherical shell is considered, and an analytical CIR is obtained for partially absorbing inner and outer surfaces with optional degradation. Multiple absorbing receivers are handled in \cite{Ferrari2022MultipleReceivers} by modeling diffusion toward several fully absorbing spheres in an unbounded medium and expressing the problem as coupled integral equations. A more general bounded model with mixed absorbing and reflecting patches on circular and spherical boundaries is presented in \cite{Chen2025HeterogeneousBoundary}, where the heterogeneous surface is homogenized into an equivalent partially reactive boundary under diffusion-only transport. On the received-signal side, \cite{Baydas2024ParameterEstimation} considers a spherical transmitter–receiver pair with partial absorption in an unbounded diffusive channel and fits a low-dimensional analytical CIR model to simulations. Finally, \cite{Zhao2025PoissonReception} is based on a 1D diffusion channel with a Poisson reception process and uses this setting to study identification-style communication. For duct-like and intrabody scenarios, \cite{Yue2024BioIoT} models micro-circulation for bio-Internet-of-Things applications using a laminar Poiseuille flow network of arterioles, capillaries, and venules, where large dimensionless numbers justify advection-dominated transport and flow-driven CIRs are derived for each vessel segment, without explicitly modeling localized reactive receiver patches. Data-driven approaches complement these physical models. In \cite{TorresGomez2023Explainability}, neural detectors are trained on synthetic CIRs and experimental traces, and explainability tools indicate that the learned networks effectively implement intuitive decision rules similar to classical peak- and edge-based detectors, while their behavior can still be interpreted using standard signal-to-noise–based BER concepts. In \cite{Ozdemir2021CaptureProbabilities}, the aim is to approximate the CIR itself for complex 2D diffusion-based topologies: transmitter and receiver layouts and reflecting obstacles are encoded as images and passed to a convolutional and fully connected neural network, which predicts the CIR shape and peak and can be superposed to approximate multi-transmitter channels with low normalized error compared to Monte-Carlo simulations.

Overall, these works demonstrate substantial progress in modeling molecular channels, but they also rely on simplifying assumptions. Analytical studies often use unbounded or highly idealized geometries and omit either flow or detailed surface reactions, while data-driven surrogates are trained on these simplified channels or focus on detection rather than physically faithful CIR prediction. These assumptions make the problems tractable, but in realistic ducts with laminar flow and reactive surfaces (e.g., in blood vessels or microfluidic channels) they can cause mismatches between the predicted channel behavior and the underlying transport dynamics and require recomputing the channel response whenever the physical parameters change. In this work, we develop a 3D duct model that jointly captures diffusion, laminar flow, and surface reaction at a finite ring-shaped wall receiver and use it to generate a large library of CIRs over a wide range of realistic channel conditions. On top of this model, we train a deep learning surrogate that maps physical parameters of the duct, the flow, and the surface chemistry directly to the full temporal channel response at low computational cost. The resulting framework serves as a benchmark channel model for evaluating detectors, equalizers, and coding schemes for reactive diffusion–advection links, and as a “ground-truth” emulator against which future analytical approximations and data-driven surrogates can be compared.

\section{system model}
\begin{figure}
	\begin{center}
		\includegraphics[width=3.3 in]{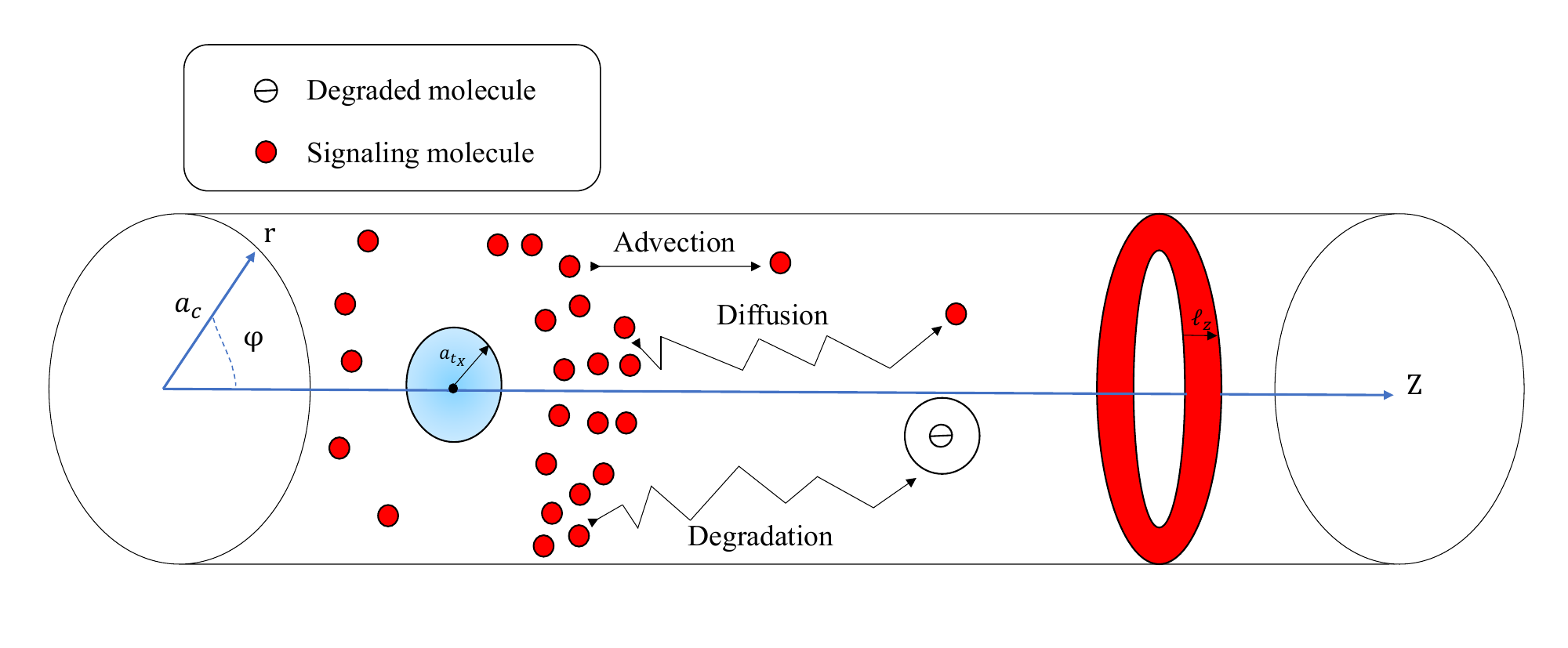}
		\caption{System geometry showing diffusion, advection, and degradation in a cylindrical duct. 
The transmitter is centered on the duct axis with spherical radius $a_{\mathrm{tx}}$, and a ring-shaped receiver of width ${\ell}_{z}$ 
is located on the wall at radius $a_{c}$.}

		\label{fig:system}
	\end{center}
\end{figure}

We consider a MC link inside a cylindrical duct as illustrated in Fig.~\ref{fig:system}, where signaling molecules propagate under diffusion, background flow, and first-order degradation. The transmitter is located at the center of the duct’s cross-section, and molecules spread throughout the fluid as they travel downstream. The duct wall is reflective everywhere except for a narrow ring-shaped sensing region that serves as the receiver. Molecules that reach this ring can bind to surface receptors and generate the measurable output signal, while all other molecules either continue propagating or undergo degradation within the channel.

\subsection{Transmitter}

We consider a finite-volume transmitter positioned on the duct axis at 
$ \mathbf{r}_{\mathrm{tx}} = (0,0,z_{\mathrm{tx}}) $, which releases $N_{0}$ identical signaling molecules of type $A$ at time $t = 0$. The transmitter occupies a small spherical region of radius $a_{\mathrm{tx}} \ll a_{c}$, where $a_{c}$ is the duct radius. At the moment of release, the molecules are assumed to be uniformly distributed within the transmitter volume, leading to the following initial concentration field \cite{Jamali2019Tutorial}:
\begin{equation}
c(\mathbf{r},0) = \frac{N_{0}}{V_{\mathrm{tx}}}\,
\Pi\!\left(\frac{\lVert \mathbf{r}-\mathbf{r}_{\mathrm{tx}} \rVert}{a_{\mathrm{tx}}}\right),
\end{equation}
where $\Pi(\cdot)$ is the rectangular function and $V_{\mathrm{tx}}$ denotes the transmitter volume 
(for a spherical transmitter, $V_{\mathrm{tx}} = \tfrac{4}{3}\pi a_{\mathrm{tx}}^{3}$). The transmitter surface is modeled as non-reactive and reflective and corresponds to the Neumann boundary condition as: 

\begin{equation}
-D \nabla c \cdot \mathbf{n} = 0,
\qquad
\mathbf{r} \in \partial\Omega_{\mathrm{tx}}.
\end{equation}

where $D$ denotes the molecular diffusivity, $\nabla c$ is the spatial concentration gradient, $\mathbf{n}$ represents the outward unit normal vector on the transmitter surface $\partial\Omega_{\mathrm{tx}}$.

\subsection{Channel Model}

We consider a straight circular duct (e.g., a microfluidic or vascular analogue) of radius $a_{c}$ and length $L$, aligned with the $z$-axis. Cylindrical coordinates $(\rho,\varphi,z)$ are used, with $0 \le \rho \le a_{c}$, $-\pi < \varphi \le \pi$, and $z \in \mathbb{R}$.

\textbf{1) Advection:}
The fluid inside the duct is assumed to be incompressible and Newtonian, with a steady, fully developed laminar Poiseuille flow directed along the $+z$ axis. The velocity profile is given by \cite{Jamali2019Tutorial}:
\begin{equation}
v_{z}(\rho) = 2\bar{v}\!\left(1 - \frac{\rho^{2}}{a_{c}^{2}}\right).
\end{equation}
The axial velocity $v_{z}(\rho)$ and the average speed $\bar{v}$ describe laminar flow in the duct. Molecules near the center move faster, while those near the wall slow down due to viscous drag. This velocity difference causes longitudinal spreading and shapes the temporal concentration profile.

\textbf{2) Diffusion:}
Molecular diffusion arises from the random thermal motion of molecules and redistributes concentration within the duct, allowing molecules to explore all radial positions. It smooths the sharp gradients formed after release and provides the transverse mixing needed for molecules not aligned with the transmitter. In the confined duct geometry, the balance between diffusion and advection determines how the plume spreads longitudinally, from weakly dispersive to fully mixed regimes \cite{Jamali2019Tutorial}.

\textbf{3) Reaction:}
In addition to physical transport, signaling molecules may undergo first-order degradation while moving through the fluid, modeled as $A \xrightarrow{\kappa} \varnothing$, where $A$ is the signaling species and $\varnothing$ denotes an inert, non-signaling state. This reaction captures natural molecular loss due to hydrolysis or enzymatic activity; when $\kappa = 0$ no degradation occurs, while larger $\kappa$ leads to faster decay and reduced molecular availability downstream \cite{Jamali2019Tutorial}.

\subsection{Receiver and Received Signal}

At the downstream end of the duct, we consider a finite annular (ring-shaped) receiver located on the inner wall. The receiver occupies a narrow axial segment centered at $z = z_{\mathrm{rx}}$ and spans the full circumference at a fixed radial position $\rho = a_{c}$. This configuration is defined as $
S_{\mathrm{rx}} 
= 
\left\{
(\rho,\varphi,z):
\ \rho = a_{c},\ 
z_{\mathrm{rx}} - \frac{\ell_{z}}{2} \le z \le z_{\mathrm{rx}} + \frac{\ell_{z}}{2},\ 
0 \le \varphi < 2\pi
\right\},$ where $\ell_{z}$ is the axial length of the reactive patch. This annular geometry is both axisymmetric and experimentally realizable, modeling a localized receptor band on a capillary or microfluidic channel wall.

\textbf{1) Receptor Dynamics:}
The receiver surface carries biochemical receptors that selectively bind to the signaling molecules $A$. Binding occurs reversibly according to \cite{Jamali2019Tutorial}:

\begin{equation}
A + B \;\xrightleftharpoons[k_{r}]{k_{f}}\; C.
\end{equation}

where $B$ denotes the surface density of unbound receptors, $C$ denotes the bound receptor–ligand complexes, $k_{f}$ is the forward binding rate, and  
$k_{r}$ is the reverse unbinding rate. The total surface receptor density $B_{\mathrm{tot}}$ is constant, such that $B + C = B_{\mathrm{tot}}$. The surface interaction between diffusing molecules and receptors is described by a mixed Robin-type boundary condition on $S_{\mathrm{rx}}$ \cite{Sun2024FiniteReceptors}:

\begin{equation}
-D\nabla c \cdot \mathbf{n}
=
k_{f}\, c(\mathbf{r},t)\big(B_{\mathrm{tot}} - C(\mathbf{r},t)\big)
-
k_{r}\, C(\mathbf{r},t),\\
\qquad \mathbf{r} \in S_{\mathrm{rx}}.
\label{eq:a}
\end{equation}

where the binding ODE for C is defined as \cite{Jamali2019Tutorial}:
\begin{equation}
\frac{\partial C}{\partial t}
= k_{f}\, c \left( B_{\mathrm{tot}} - C \right)
- k_{r} C, \qquad \mathbf{r} \in S_{\mathrm{rx}} .
\label{eq:b}
\end{equation}

\textbf{2) Channel Impulse Response:}
The measured signal at the receiver is proportional to the instantaneous \textit{binding flux} of molecules forming complexes on $S_{\mathrm{rx}}$. Accordingly, the measured rate signal is defined as:

\begin{equation}
y(t) = \int_{S_{\mathrm{rx}}} k_{f}\, c(\mathbf{r},t)\big(B_{\mathrm{tot}} - C(\mathbf{r},t)\big)\,\mathrm{d}S.
\label{eq:c}
\end{equation}

which represents the total number of new bindings per unit time across the receptor band. For analytical characterization, the CIR, $h(t)$ is defined as the expected rate of binding events per released molecule as:

\begin{equation}
h(t) = \frac{y(t)}{N_{0}}.
\label{eq:d}
\end{equation}

 To obtain the measurable signal $y(t)$ and consequently the CIR, one must solve the coupled advection–diffusion–reaction system governing the concentration of signaling molecules inside the duct and their interactions at the receiver surface. The spatiotemporal concentration of type-$A$ molecules, $c(\rho,\varphi,z,t)$, evolves under the combined effects of diffusion, advection, and chemical degradation as:

\begin{equation}
\frac{\partial c}{\partial t}
=
D\left[
\frac{1}{\rho}\frac{\partial}{\partial \rho}
\left(\rho \frac{\partial c}{\partial \rho}\right)
+
\frac{1}{\rho^{2}}\frac{\partial^{2}c}{\partial \varphi^{2}}
+
\frac{\partial^{2}c}{\partial z^{2}}
\right]
-
v_{z}(\rho)\frac{\partial c}{\partial z}
-
\kappa c,
\label{eq:e}
\end{equation}

This equation is valid within the cylindrical domain. Also, the upstream end of the duct is closed to incoming diffusion, while the downstream outlet allows molecules to leave the domain naturally with the flow. These conditions prevent artificial accumulation or loss at the domain limits as:

\begin{equation}
\frac{\partial c}{\partial z} = 0,
\qquad
\text{at } z = z_{\min},
\end{equation}

\begin{equation}
- D\frac{\partial c}{\partial z} + v_{z}c = 0,
\qquad
\text{at } z = z_{\max}.
\end{equation}

\subsection{Computational Challenge}
Solving Eqs.~\ref{eq:a}--\ref{eq:e} is computationally intensive and has no analytical solution; while numerical solvers can approximate the result, several factors make this costly. The problem involves strong coupling between the 3D bulk concentration field and nonlinear surface receptor ODEs, requiring simultaneous PDE--ODE solution. Its multi-scale behavior where diffusion, advection, and reaction operate on very different spatial and temporal scales---demands fine spatial meshes and very small time steps. Moreover, stiffness and high dimensionality arise because fast surface kinetics coupled with slow convection produce stiff equations, and the 3D cylindrical geometry yields large system matrices that make implicit solvers expensive. Consequently, analytical solutions are unavailable, and computing \(c(r,t)\) and \(h(t)\) numerically is costly, motivating fast surrogate models instead of repeated PDE solves \cite{Jamali2019Tutorial}.

%%%%%%%%%%%%%%%%%%%%%%%%%%%%%%%%%%%%%%%%%%%%%%%%%%%%%%%%%%%%%%%%
%%%%%%%%%%%%%%%%%%%%%%%%%%%%%%%%%%%%%%%%%%%%%%%%%%%%%%%%%%%%%%%%
%
\section{Deep Learning–Based Neural Surrogate for Fast CIR Prediction}
\subsection{Objectives}

To enable fast evaluation across the high-dimensional parameter space, we develop a deep learning–based surrogate that predicts the CIR directly from the system parameters without running the full PDE solver. The surrogate learns the mapping from the physical parameters to the normalized CIR defined in following subsections, where the measurable quantity is the forward binding rate on the reactive ring. Once trained, the model provides CIR predictions several orders of magnitude faster than numerical simulation, enabling real-time inference, large-scale design-space exploration, and efficient link-level evaluation \cite{Li2020FNO}.

\begin{figure*}[tbp]
\centering

\subfloat[]{%
    \includegraphics[width=2.30in]{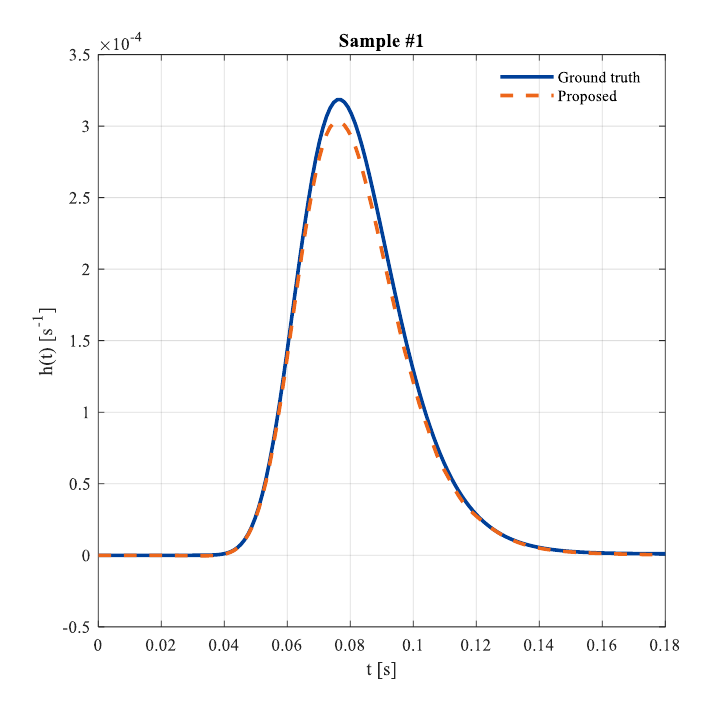}%
    \label{cf1}%
}
\hfill
\subfloat[]{%
    \includegraphics[width=2.30in]{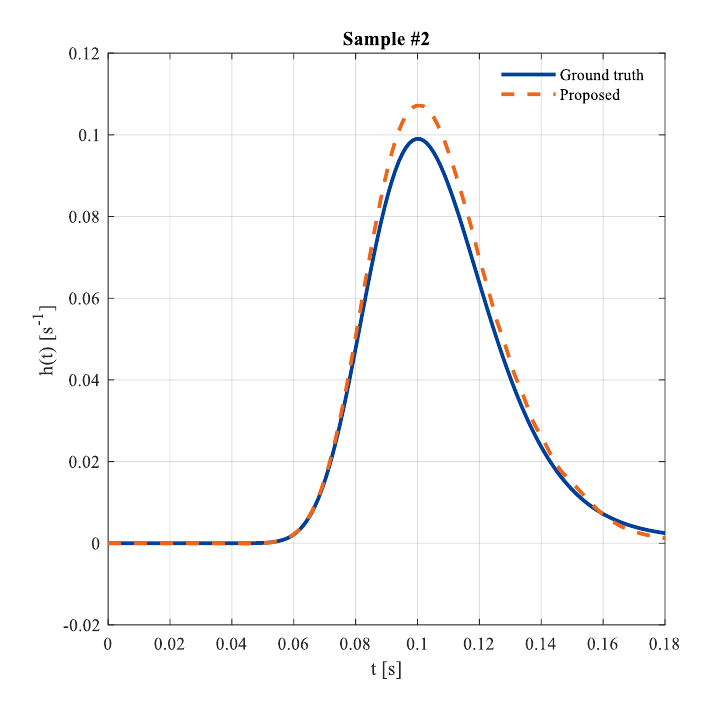}%
    \label{cf2}%
}
\hfill
\subfloat[]{%
    \includegraphics[width=2.30in]{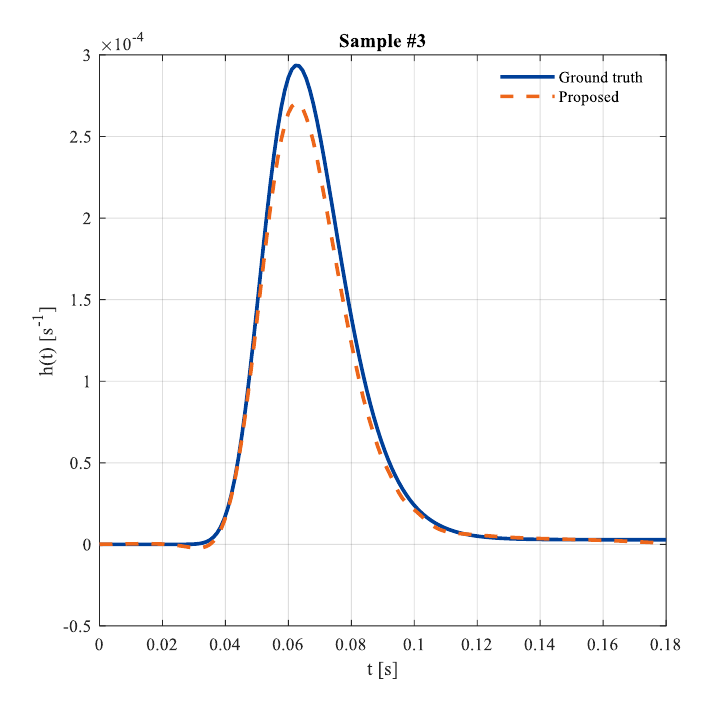}%
    \label{cf3}%
}

\caption{Comparison between ground-truth CIRs obtained from the numerical PDE solver (solid blue) and the proposed surrogate model (dashed orange) for three randomly selected test channels: (a) Sample~\#1, (b) Sample~\#2, and (c) Sample~\#3.}

\label{cf}
\end{figure*}

\subsection{Problem Formulation (Inputs \& Outputs)}

Let $
\mathbf{p}
=
\big[\, D,\ \bar{v},\ \kappa,\ k_{f},\ k_{r},\ B_{\mathrm{tot}},\ z_{\mathrm{rx}},\ \ell_{z} \,\big]^{\top},
$ denote the vector of physical and geometric parameters that vary across the dataset. For each parameter vector $\mathbf{p}$, the ground-truth CIR is the discretized signal $
\mathbf{h}(\mathbf{p})
=
\big[\, h(t_{1};\mathbf{p}),\ \ldots,\ h(t_{N_{s}};\mathbf{p}) \,\big]^{\top},$ obtained by solving the coupled PDE with nonlinear receptor kinetics, evaluated on the uniform temporal grid $\{ t_{\ell} \}_{\ell=1}^{N_{s}}$, with $N_{s}$ being the total number of time steps in the discretized CIR. The goal is to train a deep neural surrogate $\hat{\mathbf{h}}(\mathbf{p})$ such that:
\begin{equation}
\hat{\mathbf{h}}(\mathbf{p}) \;\approx\; \mathbf{h}(\mathbf{p})
\qquad \text{for all }\mathbf{p}\text{ in the design domain},
\end{equation}

Beyond the raw input parameters $\mathbf{p}$, we construct additional quantities that summarize the dominant transport and reaction scalings:

\[
\begin{aligned}
\mathrm{Pe} &= \frac{\bar{v} a_{c}}{D}, 
&\qquad
\mathrm{Da} &= \frac{k_{f} B_{\mathrm{tot}} a_{c}}{D}, \\[6pt]
t_{\mathrm{diff}} &= \frac{a_{c}^{2}}{4D},
&\qquad
t_{\mathrm{adv}} &= \frac{z_{\mathrm{rx}}}{\bar{v}}, \\[6pt]
\zeta &= \frac{t_{\mathrm{adv}}}{t_{\mathrm{diff}}},
&\qquad
A_{\mathrm{patch}} &= 2\pi a_{c}\ell_{z}, \\[6pt]
R_{\mathrm{cap}} &= B_{\mathrm{tot}} A_{\mathrm{patch}},
&\qquad
k_{\mathrm{dim}} &= \kappa\, t_{\mathrm{adv}}.
\end{aligned}
\]

These eight derived features encode advection–diffusion balance (Pe), bulk-to-surface coupling strength (Da), characteristic diffusion and advection time scales ($t_{\mathrm{diff}},\, t_{\mathrm{adv}},\, \zeta$), reactive-patch geometry and receptor capacity ($A_{\mathrm{patch}},\, R_{\mathrm{cap}}$), and bulk decay over one advective time ($k_{\mathrm{dim}}$).

We concatenate the 8 raw parameters and these 8 derived features, giving a 16-dimensional feature vector $X \in \mathbb{R}^{16}$. This modest expansion keeps the input space low-dimensional while still exposing the main physical balances to the network.

To reduce dynamic range, we apply a log-transform to all strictly positive entries in our feature vector $X$ and then standardize all features via:
\begin{equation}
X^{(z)} = \frac{X - \mu_{X}}{\sigma_{X}},
\end{equation}

with $\mu_{X}$ and $\sigma_{X}$ computed from the training partition.

\textbf{1) Time–Warped Unit-Shape Factorization:}
The CIRs generated by the full PDE–ODE solver differ substantially in peak magnitude, arrival time, and temporal dispersion. To stabilize learning and reduce the complexity the network must model, we factor each waveform into physically interpretable scale–shift–shape components. Let each CIR be sampled on a fixed temporal grid $t_{\ell}$, producing the discrete vector $\mathbf{h} = [h_{1},\ldots,h_{N_{s}}]^{\top}, \qquad h_{\ell} = h(t_{\ell})$, we extract three scalar summary parameters: $A = \max_{\ell}\, h_{\ell}, \qquad t_{p} = t_{\arg\max_{\ell} h_{\ell}},$ $
w
=
\sqrt{
\frac{
\sum_{\ell=1}^{N_{s}} h_{\ell}(t_{\ell} - t_{p})^{2}
}{
\sum_{\ell=1}^{N_{s}} h_{\ell}
}
}
$ \cite{Ramsay2005FDA}.

Here $A$ is the peak amplitude, $t_{p}$ is the arrival time of the peak, and $w$ is an energy-weighted rms width capturing temporal spread. We normalize both time and amplitude via \cite{Ramsay2005FDA}:

\begin{equation}
\tau_{\ell} = \frac{t_{\ell} - t_{p}}{w},
\end{equation}
\begin{equation}
y_{n}(\tau_{\ell}) = \frac{h_{\ell}}{A}.
\end{equation}

The normalized waveform $y_{n}(\tau)$ is then resampled onto a fixed uniformly spaced $\tau$-grid, shared across all samples. During training, the neural network predicts both the unit-shape $y_{n}(\tau)$ and the triplet $(A,\, t_{p},\, w)$. At inference time, the physical CIR is reconstructed via:
\begin{equation}
h(t) = A\, y_{n}\!\left( \frac{t - t_{p}}{w} \right),
\end{equation}

Providing a stable and compact representation across widely different regimes.

\textbf{2) Weighted PCA for Shape Basis Extraction:}
After time normalization, each CIR is represented by a dimensionless unit shape $y_{n}(\tau)$. We assemble these into the matrix $H_{n} \in \mathbb{R}^{N_{s}\times N_{L}}$, where each column corresponds to one normalized shaped $y_{n}$. To emphasize the peak-neighborhood where most transport and reaction-related differences occur we perform PCA using a time dependent weight function \cite{Ramsay2005FDA}:

\begin{equation}
w(\tau_{\ell})
=
\exp\!\left[
-\left(\frac{\tau_{\ell}}{\sigma_{w}}\right)^{2}
\right],
\end{equation}
\begin{equation}
\sum_{\ell=1}^{N_{s}} w(\tau_{\ell}) = N_{s},
\end{equation}

\begin{figure}
	\begin{center}
		\includegraphics[width=3.3 in]{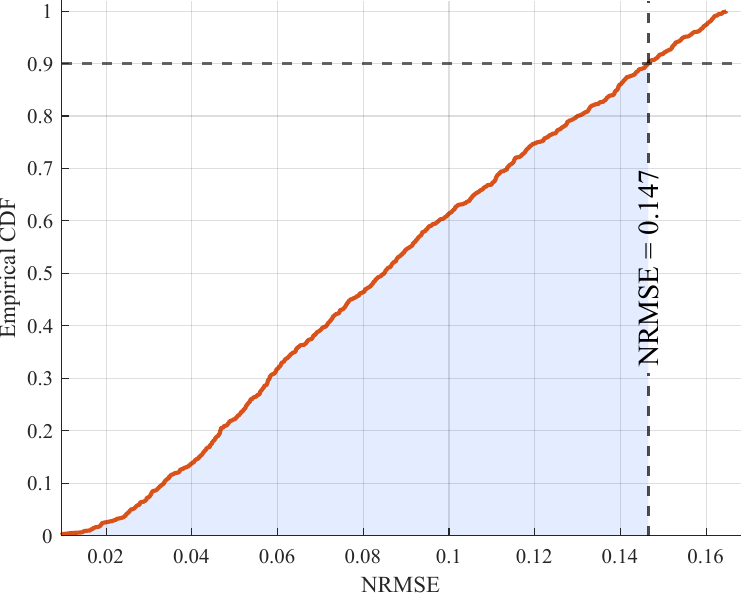}
		\caption{Empirical CDF of the NRMSE between the surrogate predictions and numerical CIRs over the test set.}

		\label{fig:c}
	\end{center}
\end{figure}

where $\sigma_{w}$ controls the temporal width of the weighting window. Let $\mu(\tau_{\ell})$ be the weighted mean, $\{ \phi_{k}(\tau_{\ell}) \}_{k=1}^{K}$ the leading weighted principal components, and $\mathbf{c} = [\, c_{1}, \ldots, c_{K} \,]^{\top}$ the coefficients. Each unit shape admits the approximation \cite{Ramsay2005FDA}:

\begin{equation}
y_{n}(\tau_{\ell})
\;\approx\;
\mu(\tau_{\ell})
+
\sum_{k=1}^{K} c_{k}\, \phi_{k}(\tau_{\ell}).
\end{equation}

We choose the smallest $K$ that captures at least 99.5\% of the weighted variance. Across our dataset this yields $20 \le K \le 28$, depending on the parameter distribution, providing a compact basis that avoids fitting noise near the peak.

\textbf{3) Multi-Output Targets and Normalization:}
The surrogate network outputs an interpretable set of parameters consisting of the PCA shape coefficients together with the physical summary scalars $(A,\, t_{p},\, w)$. We collect these into the target vector:
\begin{equation}
\mathbf{y}_{\mathrm{raw}}
=
[\, c_{1},\ldots,c_{K},\ \log A,\ t_{p},\ \log w \,]^{\top}
\in \mathbb{R}^{m},
\qquad
m = K + 3.
\end{equation}

Log-scaling is applied to $A$ and $w$ to reduce dynamic range, while $t_{p}$ is kept in linear units. Each component of $\mathbf{y}_{\mathrm{raw}}$ is standardized using the training set statistics:

\begin{equation}
\tilde{\mathbf{y}}
=
(\mathbf{y}_{\mathrm{raw}} - \mu_{Y}) \oslash \sigma_{Y},
\end{equation}

where $\mu_{Y}$ and $\sigma_{Y}$ denote the empirical mean and standard deviation of each output dimension, and $\oslash$ denotes elementwise division. To ensure that the loss emphasizes physically informative quantities particularly the peak height, arrival time, and temporal width we apply a fixed elementwise weight vector \cite{Ramsay2005FDA}:

\begin{equation}
\mathbf{W}
=
[\, \underbrace{1,\ldots,1}_{K\ \text{shape coeffs}},\ 4.0,\ 5.0,\ 3.0 \,],
\end{equation}

The reweighted target is:

\begin{equation}
\tilde{\mathbf{y}}^{(W)} = \mathbf{W} \odot \tilde{\mathbf{y}},
\end{equation}

where $\odot$ denotes elementwise multiplication.

\textbf{4) Network Architecture:}
The surrogate model is a multilayer perceptron (MLP) that maps the standardized input  features $\tilde{\mathbf{x}}$ to the reweighted target $\tilde{\mathbf{y}}^{(W)}$. The network consists of fully connected layers of sizes $192$, $192$, $96$, and $m = K + 3$, with LayerNorm applied after the first two linear layers to stabilize the pre-activation statistics. Each hidden layer uses ReLU activations, and a Dropout rate of $0.1$ is applied after the first layer for regularization, while no activation is used after the final output layer. The network parameters $\theta$ are learned by minimizing a weighted mean-squared error between the predicted outputs and the standardized targets. For a minibatch $B$, the loss is:

\begin{equation}
\mathcal{L}(\theta)
=
\frac{1}{|B|}
\sum_{i \in B}
\left\|
\tilde{\mathbf{y}}^{(W)}_{i}
-
f_{\theta}(\tilde{\mathbf{x}}_{i})
\right\|_{2}^{2}
+
\lambda \|\theta\|_{2}^{2},
\end{equation}

where $f_{\theta}(\cdot)$ is the MLP introduced above, $\tilde{\mathbf{x}}_{i}$ is the standardized feature vector of sample $i$, $\tilde{\mathbf{y}}^{(W)}_{i}$ is the corresponding reweighted output target, and $\lambda$ is the weight-decay coefficient.

Given the ensemble prediction $\hat{\mathbf{y}}^{(W)}$, we first undo the output reweighting and standardization:

\begin{equation}
\tilde{\mathbf{y}} = \hat{\mathbf{y}}^{(W)} \oslash \mathbf{W},
\end{equation}
\begin{equation}
\hat{\mathbf{y}} = \tilde{\mathbf{y}} \odot \sigma_{Y} + \mu_{Y}.
\end{equation}

Using the PCA basis, the normalized shape is reconstructed as \cite{Ramsay2005FDA}:

\begin{equation}
\hat{y}_{n}(\tau_{\ell})
=
\mu(\tau_{\ell})
+
\sum_{k=1}^{K}
\hat{c}_{k}\,\phi_{k}(\tau_{\ell}).
\end{equation}

Finally, the physical CIR is obtained by reversing the time-warp. The normalized shape $\hat{y}_{n}(\tau)$ is interpolated to evaluate at $\tau = (t - \hat{t}_{p}) / \hat{w}$, yielding:

\begin{equation}
\hat{h}(t)
=
\hat{A}\,
\hat{y}_{n}\!\left(
\frac{t - \hat{t}_{p}}{\hat{w}}
\right).
\end{equation}

\section{SIMULATION RESULTS AND DISCUSSION}

Unless stated otherwise, all results are based on a dataset of $N = 10k$ CIRs generated by numerically solving the advection diffusion reaction system in Eq.~\ref{eq:d}. Also, 
$
D \in [\,5\times 10^{-10},\ 2\times 10^{-9}\,]\ \mathrm{m^{2}/s},
$
$
\bar{v} \in [\,1\times 10^{-3},\ 3\times 10^{-3}\,]\ \mathrm{m/s},
$
$
\kappa \in [\,0,\ 1\,]\ \mathrm{s^{-1}},
$
$
k_{f} \in [\,1\times 10^{-6},\ 5\times 10^{-6}\,]\ \mathrm{m^{3}/(molecule\cdot s)},
$
$
k_{r} \in [\,5\times 10^{-2},\ 5\times 10^{-1}\,]\ \mathrm{s^{-1}},
$
$
B_{\mathrm{tot}} \in [\,10^{15},\ 10^{16}\,]\ \mathrm{molecules/m^{2}},
$
$
z_{\mathrm{rx}} \in [\,150,\ 300\,]\ \mu\mathrm{m},
$
$
\ell_{z} \in [\,10,\ 40\,]\ \mu\mathrm{m}
$ \cite{Jamali2019Tutorial}. The duct radius $a_{c} = 60\,\mu\mathrm{m}$, transmitter radius $a_{\mathrm{tx}} = 10\,\mu\mathrm{m}$, and injected molecule count $N_{0} = 1$ are kept fixed for all realizations. The resulting $10k$ samples are randomly partitioned into $N_{\mathrm{tr}} = 7k$ training, $N_{\mathrm{val}} = 1.5k$ validation, and $N_{\mathrm{te}} = 1.5k$ test waveforms. The neural surrogate is trained on the training set, hyperparameters are selected on the validation set, and all performance metrics are reported on the held-out test set.

\begin{table}[t]
\centering
\caption{Correlation Between Physical Parameters and NRMSE}
\begin{tabular}{lcc}
\hline
\textbf{ParamName} & \textbf{ParamMean} & \textbf{CorrWithNRMSE} \\
\hline
$D$        & $1.1025\times10^{-9}$   & $-0.083366$ \\
$\bar{v}$  & $2.0850\times10^{-3}$    & $-0.12773$  \\
$\kappa$   & $4.9295\times10^{-1}$    & $0.031381$  \\
$k_{\text{f}}$   & $2.4548\times10^{-6}$ & $-0.016943$ \\
$k_{\text{r}}$  & $1.9879\times10^{-1}$ & $-0.0079414$ \\
$B_{\mathrm{tot}}$ & $3.8799\times10^{15}$ & $0.064297$ \\
$z_{\mathrm{rx}}$ & $2.2325\times10^{-4}$ & $0.03289$ \\
${\ell}_{\mathrm{z}}$ & $2.6066\times10^{-5}$ & $-0.059911$ \\
\hline
\end{tabular}
\label{tab:corr_params}
\end{table}

Fig.~\ref{cf} compares the ground-truth CIRs obtained from the numerical PDE solver (solid blue) with the outputs of the proposed surrogate model (dashed orange) for three randomly selected samples from the independent test set. For all three cases, the predicted responses follow the reference CIRs closely over the whole time interval, reproducing the rise, peak position, and subsequent decay without visible artefacts. Small deviations appear mainly around the peak region, but the overall shape and amplitude are well preserved. These representative examples suggest that the surrogate can approximate the CIR with good accuracy for different channel conditions; this observation is quantified in Fig.~\ref{fig:c}, which shows the empirical cumulative distribution function (CDF) of the normalized root mean square error (NRMSE) over the test set. Each point on the curve corresponds to one test CIR, where the NRMSE measures the root–mean–square deviation between the predicted and reference CIR, normalized by the energy of the reference response. The CDF increases smoothly from very small errors up to about 0.16, indicating that there are no catastrophic outliers in the evaluated range. The dashed lines highlight that 90\% of the test samples exhibit an NRMSE below 0.147, i.e., for the vast majority of channels the surrogate reproduces the ground-truth CIR with a relative error of less than roughly 15\%. This distribution-based view complements the qualitative overlays in Fig.~\ref{cf} and shows that the proposed model achieves consistently low approximation error across a wide variety of channel conditions, rather than only for a few isolated examples.

Table~I reports the Pearson correlation between the NRMSE and each physical parameter on the test set, together with the corresponding mean values. All coefficients lie in $[-0.13,\,0.07]$, indicating only weak linear dependence on any single parameter. The strongest effects occur for the average flow velocity $\bar{v}$ ($\rho \approx -0.13$) and diffusion coefficient $D$ ($\rho \approx -0.08$), while $\kappa$, $k_{\mathrm{f}}$, $k_{\mathrm{r}}$, $B_{\mathrm{tot}}$, $z_{\mathrm{rx}}$, and ${\ell}_{\mathrm{z}}$ show even smaller correlations. Thus, the surrogate does not systematically deteriorate for particular transport, reaction, or geometric regimes, and maintains a nearly uniform level of accuracy across the explored channel parameters.

\section{Conclusion}
This paper presented a deep-learning surrogate for a reactive diffusion–advection channel in a cylindrical duct, trained on physics-based PDE-generated CIRs. The surrogate maps channel parameters directly to the CIR and closely matches the numerical ground truth across diverse conditions, reproducing the main features of the response. The empirical CDF of the NRMSE over an independent test set shows that 90\% of channels are approximated with a normalized error below about 0.15, while correlation analysis reveals only weak dependence of the residual error on transport, reaction, and geometric parameters, indicating fairly uniform accuracy over the explored space. These results demonstrate that the proposed surrogate can serve as an accurate and computationally efficient replacement for repeated PDE solves, enabling fast performance evaluation and design of MC systems.
%%%%%%%%%%%%%%%%%%%%%%%%%%%%%%%%%%%%%%%%%%%%%%%%%%%%%%%%%%%%%%%%
%%%%%%%%%%%%%%%%%%%%%%%%%%%%%%%%%%%%%%%%%%%%%%%%%%%%%%%%%%%%%%%% VERSUS W_Z

%%%%%%%%%%%%%%%%%%%%%%%%%%%%%%%%%%%%%%%%%%%%%%%%%%%%%%%%%%%%%%%%
%%%%%%%%%%%%%%%%%%%%%%%%%%%%%%%%%%%%%%%%%%%%%%%%%%%%%%%%%%%%%%%% 

%%%%%%%%%%%%%%%%%%%%%%%%%%%%%%%%%%%%%%%%%%%%%%%%%%%%%%%%%%%%%%%%
%%%%%%%%%%%%%%%%%%%%%%%%%%%%%%%%%%%%%%%%%%%%%%%%%%%%%%%%%%%%%%%% VERSUS W_Z

%%%%%%%%%%%%%%%%%%%%%%%%%%%%%%%%%%%%%%%%%%%%%%%%%%%%%%%%%%%%%%
%%%%%%%%%%%%%%%%%%%%%%%%%%%%%%%%%%%%%%%%%%%%%%%%%%%%%%%%%%%%%%
\bibliographystyle{IEEEtran}
\bibliography{myref}

\end{document}